\documentclass[fleqn,usenatbib]{mnras}

\usepackage{newtxtext,newtxmath}

\usepackage[T1]{fontenc}
\usepackage{ae,aecompl}

\usepackage{graphicx}	
\usepackage{amsmath}	
\usepackage{amssymb}	

\newcommand{\degree}{$^{\circ}$}
\defcitealias{leroy}{Leroy et al. (submitted)}

\title{A Comparison of Stellar and Gas-Phase Chemical Abundances in Dusty Early-Type Galaxies}

\author[Griffith, Martini, \& Conroy]{Emily Griffith$^{1}$, Paul Martini$^{1,2}$, and Charlie Conroy$^{3}$
\\
$^{1}$Department of Astronomy, The Ohio State University, 140 West 18th Avenue, Columbus, OH 43210, USA\\
$^{2}$Center for Cosmology and Astro-Particle Physics, The Ohio State University, 191 West Woodfuff Avenue, Columbus OH, 43210, USA\\
$^{3}$Department of Astronomy, Harvard University, Cambridge, MA 02138, USA\\
}
\date{Last updated \today}

\pubyear{2018}

\begin{document}
\label{firstpage}
\pagerange{\pageref{firstpage}--\pageref{lastpage}}
\maketitle

\begin{abstract}
While we observe a large amount of cold interstellar gas and dust in a subset of the early-type galaxy (ETG) population, the source of this material remains unclear. The two main, competing scenarios are external accretion of lower mass, gas-rich dwarfs and internal production from stellar mass loss and/or cooling from the hot interstellar medium. We test these hypotheses with measurements of the stellar and nebular metallicities of three ETGs (NGC 2768, NGC 3245, and NGC 4694) from new long-slit, high signal-to-noise ratio spectroscopy from the Multi-Object Double Spectographs on the Large Binocular Telescope. These ETGs have  modest star formation rates and minimal evidence of nuclear activity. We model the stellar continuum to derive chemical abundances and measure gas-phase abundances with standard nebular diagnostics. We find that the stellar and gas-phase abundances are very similar, which supports internal production and is very inconsistent with the accretion of smaller, lower metallicity dwarfs. All three of these galaxies are also consistent with an extrapolation of the mass-metallicity relation to higher mass galaxies with lower specific star formation rates. The emission line flux ratios along the long-slit, as well as global line ratios clearly indicate that photoionization dominates and ionization by alternate sources including AGN activity, shocks, cosmic rays, dissipative magnetohydrodynamic waves, and single degenerate Type Ia supernovae progenitors do not significantly affect the line ratios.
\end{abstract}


\section{Introduction}
\label{sec:introduction}

One simplistic way to distinguish late and early-type galaxies is by their dust content and star formation. While late-type galaxies contain large amounts of dust and have on going star formation, early-type galaxies (ETGs) were thought to lack cold interstellar material and have lower star formation rates. However, infrared observations of ETGs starting in the late 1980s revealed cold, interstellar dust in a substantial subset of the population \citep{jura,knapp,goudfrooij,bregman}. Subsequent studies of ETGs with archival \textit{Hubble Space Telescope} (\textit{HST}) and \textit{Spitzer} images show that approximately 60\% contain a substantial amount of interstellar dust--on the order of $10^6\text{ M}_{\odot}$ \citep{smith12, martini13}. Though the source of this dust was unclear, the two most common scenarios are external accretion through interactions with low metallicity dwarf galaxies and internal production by evolved star mass loss and/or gas cooling from the hot phase of the interstellar medium (ISM).

Mass loss through gas and dust ejection from evolved stars \citep[e.g.][]{hofner} could explain the 10-12 $\mu$m dust signatures seen in many elliptical galaxies \citep{knapp92}, but the viability of internal production requires an equilibrium between dust production and destruction to sustain the observed dust masses. Thermal sputtering in the hot ISM \citep{draine_b} should destroy dust on a timescale of $\tau_{\text{dust}} \approx 2 \times 10^{4} - 10^{7}$ yr \citep{goudfrooij,clemens,smith12}. Given a destruction time of $\tau_{\text{dust}} = 2 \times 10^4$ yr and assuming that stars eject material at a constant rate, \citet{martini13} derived an internally produced, steady-state dust mass of $10-100 \text{ M}_{\odot}$--orders of magnitude lower than the $10^{5}$ to $10^{6.5} \text{ M}_{\odot}$ of dust detected in about half of the population. Similar work by \citet{rowland} and \citet{smith12} conclude that dust destruction timescales would need to be much longer in order to produce the observed dust masses.

The inclusion of cooling from the hot phase of the ISM in dust lifetime models could explain the observed dust properties. Assuming characteristic Milky Way dust lifetimes ($10^7-10^8$ yr) and stellar mass loss rates suggested by mid-infrared ETG emission ($0.1-1 \text{ M}_{\odot} \text{ yr}^{-1}$), mass loss from AGB stars could produce gas reservoirs of $5\times 10^3$ to $5\times 10^4 \text{ M}_{\odot}$ \citep{athey}. After entering the hot ISM, these reservoirs of dusty gas could cool through thermal collisions \citep{mathews}. The dust would then clump and collect into clouds, shielding the material and producing the observed emission. 

With the addition of cooling, internal dust production becomes a plausible dust source.  However, it does not explain the lack of dust in some ETGs. Given ETGs' similar evolution and stellar properties, we expect consistent dust masses across similar populations. Instead, only about 50-60\% of ETGs contain measurable amounts of dust \citep{tran, smith12, martini13}. The remainder show no evidence for interstellar dust. The bifurcation between dusty and dustless galaxies, rather than a continuum of dust masses, is inconsistent with internal production. These demographics suggest that stellar mass loss and gas cooling cannot be the only source of interstellar dust in ETGs, or that there is some other important deviation from their apparent self-similarity. 

Alternatively, external accretion of dwarf galaxies could explain the surplus of dust in a subset of the population. \textit{HST} images reveal that many ETGs contain chaotic and clumpy dust lanes--morphology suggestive of past merger events \citep{vanD95}. There is also significant kinematic evidence for a rich merger history from alignments between the molecular gas and the stellar populations, most recently from ATLAS$^{\text{3D}}$ \citep{bertola92, morganti, young, davis, cappellari16}. Follow up studies of some ETGs have searched for neighbors which could serve as external gas sources, often finding plausible companions \citep{duc,crocker}. Yet the high occurrence rate of dusty ETGs presents a challenge to the external origin hypothesis. \citet{martini13} concluded that if 60\% of all ETGs contain large amounts of dust and the destruction timescale is not more than $10^7 - 10^8$ years, then the merger rate between ETGs and gas-rich dwarf galaxies must be orders of magnitude larger than the predicted value \citep{stewart}. Uncertainties in merger rates and dust lifetimes could make external accretion a viable gas and dust source, but \citet{martini13} proposed that the more likely solution is a combination of external and internal processes.

We can identify which processes produce the majority of the observed dust through a comparison of stellar and gas-phase abundances \citep{martini13}. If evolved stars eject material into the ISM, we expect the gas-phase measurements to be similar to the stellar abundances. If low metallicity dwarf galaxies have been accreted, we expect a sub-stellar gas-phase oxygen abundance as the dwarf galaxies would dilute the metallicity. We also expect to find sub-stellar oxygen abundance if a combination of the two production mechanisms are important \citep{bresolin13}.  

It is not straightforward to measure the parameters of stellar population composite stellar spectra. Since the 1980's, models of spectral indices \citep{burstein,worthey} have been used to broadly determine the ages and metallicities of stellar populations. Spectral indices are informative but ignore a great deal of information. By modeling the entire spectrum it is possible to achieve greater sensitivity to trace elements \citep{conroy13} and better quantify population parameters. This approach requires high quality stellar spectral models, such as those from the MILES library \citep{sanchez06} and the extended IRTF library \citep{villaume17}. Recently \citet{conroy12} developed such a program and have studied the underlying stellar populations of ETGs and globular clusters \citep{conroy14, conroy18}, finding good agreement with results from previous Lick index fitting techniques, such as \citet{graves}. We employ similar methods to determine the stellar properties of the ETGs included in this work. 

Gas-phase metallicities can be spectroscopically determined through a variety of techniques. The direct method, often referred to as the $T_e$ method, provides the most reliable and accurate abundance diagnostic \citep{dinerstein}. In this approach, one measures the electron temperature through measurements of auroral and nebular lines. Detecting auroral lines is extremely difficult, especially at high metallicity. Strong-line methods were subsequently developed to provide oxygen abundance estimates for galaxies without detectable auroral lines \citep{alloin,pagel} and calibrated by theoretical \citep{mcgaugh,zaritsky,kd02} and empirical means \citep{pp04,marino}. Strong-line methods employ nebular emission lines, such as $[\ion{O}{ii}]\lambda3727, [\ion{O}{iii}]\lambda5007$, and $[\ion{N}{ii}]\lambda6583$, along with prominent Balmer lines. 

In this paper we present strong-line oxygen abundance measurements of three dusty early-type galaxies observed with The Multi-Object Double Spectrographs (MODS) \citep{pogge} on the Large Binocular Telescope (LBT). We also include data from similar studies of ETG abundances by \citet{athey09}, \citet{annibali}, and \citet{bresolin13} where applicable. While few studies have compared stellar and nebular abundances for the same galaxies, we note that \citet{zahid} compare stellar and nebular abundances for star-forming galaxies. After a brief description of our data in Section~\ref{sec:methods}, in Section~\ref{sec:analysis} we fit the background stellar population, determine line fluxes, and calculate oxygen abundances. In Section~\ref{sec:results} we compare the gas-phase oxygen abundance to the stellar abundance and present the mass-metallicity relationships. Section~\ref{sec:discussion} discusses the potential for internal dust production or external accretion. Finally, we summarize our results in Section~\ref{sec:summary}.

\section{Methods}
\label{sec:methods}


\citet{martini13}'s study of archival \textit{Spitzer} and \textit{HST} data found that approximately 60\% of ETGs contained $\geq 10^5 \text{ M}_{\odot}$ of dust. To probe the nature of this interstellar material, we selected three dusty ETGs for follow up spectroscopy: NGC 2768, NGC 3245, and NGC 4694. These galaxies were chosen due to their proximity, strong presence of dust lanes in \textit{HST} imaging, and weak (NGC 2768) or no evidence (NGC 3245 \& NGC 4694) of an AGN. The filamentary dust structures, \ion{H}{I} and CO kinematics, and molecular gas masses of these galaxies all suggested external accretion as the source of interstellar dust \citep{simoes, oosterloo, crocker}. 

\begin{table}
    \caption{Observing log (top) and properties (bottom) of sample ETGs.}
    \centering
    \begin{tabular}{c c c c}
        \hline
        \hline
          & NGC 2768 & NGC 3245 & NGC 4694  \\
         \hline
         Date & 01-05-2014 & 01-25-2015 & 04-29-2014\\
         Exp. Time & 3600 s & 1200 s & 900 s\\
         Slit Width & 0.6" &1.2" & 0.6"\\
         Seeing & $\sim$1.5" & $\sim$1.5" & $\sim$1.2" \\
         Avg. Air Mass & 1.18 & 1.21 & 1.24\\
         Position Angle $^{\text{a}}$ & 15\degree & 110\degree & 140\degree \\
         Parallactic Angle & -117\degree & -70\degree & -50\degree \\
         \hline
         $\log{\text{M}_{*}} (\text{M}_{\odot})$ & 11.10 & 10.75 & 10.05 \\
        $z$ $^{\text{b}}$& 0.004590 &  0.004453 &  0.00393 \\
        D (Mpc) $^{\text{c}}$ & 21.8 & 21.0 & 16.5 \\
        SFR $(\text{M}_{\odot}$ yr$^{-1})$ $^{\text{d}}$ & 0.173 & 0.141 & 0.099 \\
        sSFR (yr$^{-1}$) & $1.4\times10^{-12}$ & $2.5\times10^{-12}$ & $8.8\times10^{-12}$ \\
        \hline
    \end{tabular}
    \begin{flushleft}
        \textbf{Notes:} 
        $^{\text{a}}$ The position angle refers to the angle of the slit. \\
        $^{\text{b}}$ Redshifts are taken from \citet{redshift}. \\
        $^{\text{c}}$ Distances are taken from \citet{cappellari}. \\
        $^{\text{d}}$ Star formation rates and stellar masses are derived in Section~\ref{sec:mass}.
    \end{flushleft}
    \label{tab:obs}
\end{table}

We observed the three ETGs between January 2014 and February 2015 with MODS1 on the LBT. The spectrograph's dichroic divides incoming light into red and blue channels at $\sim5650$\AA. The blue spectrum has a lower bound of $\sim3200$\AA\ and the red an upper bound of $\sim10,000$\AA. Both channels have a resolution of $R\sim2000$ for a 0.6" slit. The long-slit was not oriented along the parallactic angle for NGC 2768 in order to measure extended emission-line regions. As these galaxies are well resolved, they did not require good seeing and thus were observed in poorer conditions. A log of our observations and the basic properties of our sample are listed in Table~\ref{tab:obs}.

A detailed discussion of the MODS data reduction pipeline can be found in \citet{berg15}. In short, we use the modsCCDRed\footnote{\url{http://www.astronomy.ohio-state.edu/MODS/Software/modsCCDRed/}} and modsIDL\footnote{\url{http://www.astronomy.ohio-state.edu/MODS/Software/modsIDL/}} suites to bias subtract, flat field, median combine, remove sky lines, and extract 1D spectra. Unless otherwise specified, all spectra and derived quantities are based on a summation of the galaxy flux in the central 24'' of the longslit. Outside of the MODS pipeline, we apply a response curve to correct for Galactic extinction. We employ the \citet{cardelli} reddening law with $R_v = 3.1$ and use dust maps from \citet{schlegel}.

\section{Analysis}
\label{sec:analysis}

\subsection{Stellar Population Modeling}
\label{sec:alf}

After reducing the data, we model our spectra with \texttt{alf}, an ``absorption line fitting'' program \citep{conroy12,conroy18}, based on the MIST iscochrones \citep{choi16}, and optical and NIR empirical stellar libraries \citep{sanchez06, villaume17}. We use \texttt{alf} to determine stellar abundances and to better measure the emission lines. In particular, stellar absorption features can hide emission lines such as $\text{H}\alpha$ and $\text{H}\beta$.   

\begin{figure*}
 \includegraphics[width=\textwidth]{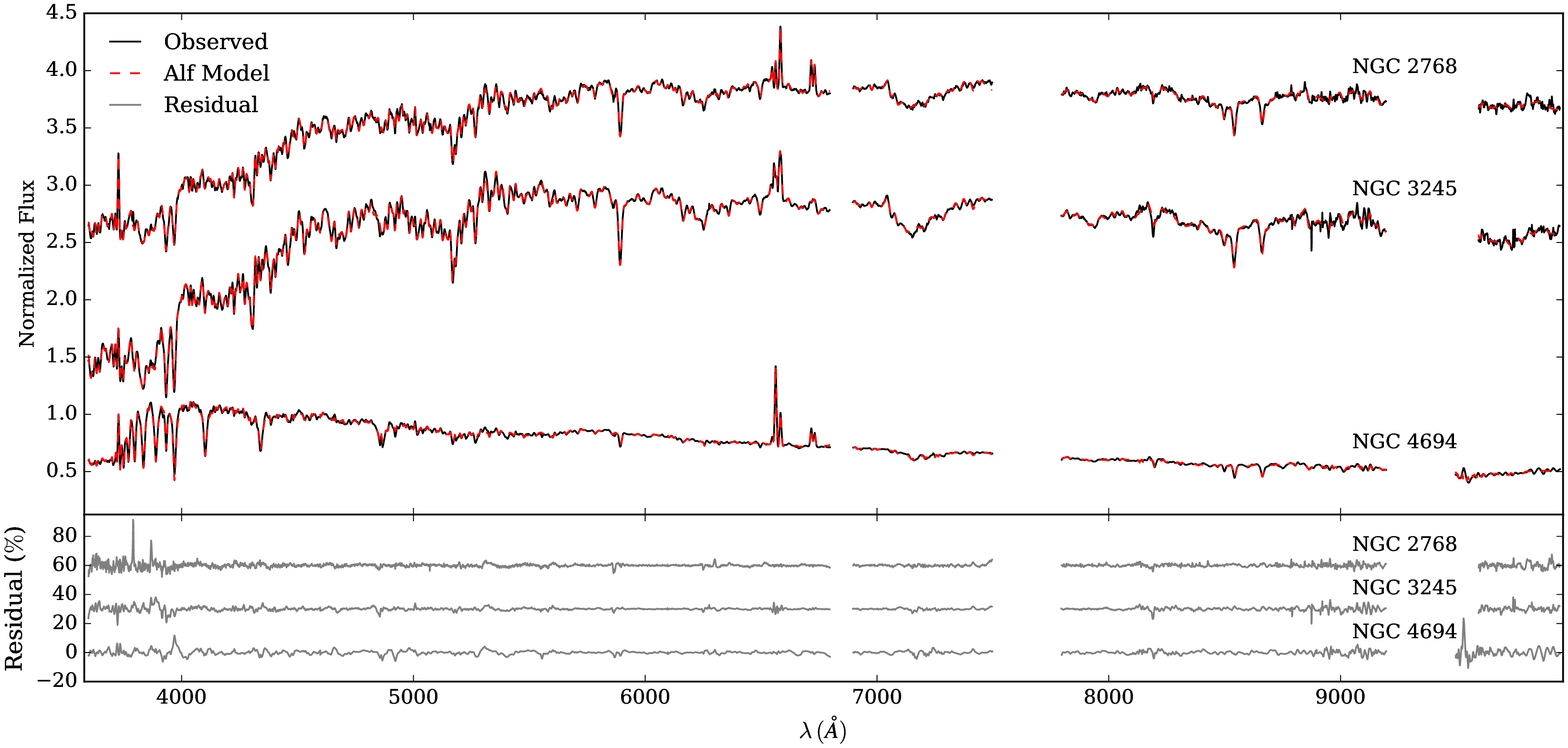}
 \caption{Top panel: MODS spectra (black line) corrected for Galactic extinction. The \texttt{alf} models of emission and absorption features are over plotted (red, dashed line). Spectra are a summation over the central 24'' along the slit. Fluxes are normalized such that the peak of [\ion{O}{ii}] $\lambda 3727 = 1$. Observed and model spectra for NGC 3245 and NGC 2768 have been shifted by 0.75 and 2.25, respectively. Telluric A and B bands, as well as a noisy region near 9400\AA, have been masked. Bottom panel: Residuals between observed and model spectra (grey line). Residuals for NGC 3245 and NGC 2768 have been shifted by 30\% and 60\%, respectively.}
 \label{fig:alf_fit}
\end{figure*}

Given some initial conditions, \texttt{alf} constrains 46 fit parameters (including stellar abundances, age, and emission line fluxes) and produces a corresponding model spectrum.  Figure~\ref{fig:alf_fit} shows the three galaxies' observed spectra, the \texttt{alf} model, and the residual between the two. Telluric features in the A and B bands have been masked, as well as a noisy region between 9200\AA\ and 9600\AA. The observed spectra shown in the figures only include a correction for Galactic extinction. Both the galaxy spectra and the models include any reddening intrinsic to the galaxies. We note the flatness of the residual spectrum--a testament to the fit quality. Residuals increase bluewards of 4000\AA\ where the density of atomic lines increases. NGC 4694's velocity dispersion \citep{weger} falls below the lower limit of \texttt{alf}'s models (100 km/s). We therefore smooth the spectrum by a Gaussian with a velocity full-width-half-maximum of 350 km/s. 

While the program reports 18 stellar abundance parameters, resolved and unresolved spectral features cause some elements to be better constrained than others. We focus on C, N, O, and Mg, which should be among the best measurements. We convert these abundances from [X/H] to [X/Fe], where [X/Y] $= \log(\text{X}/\text{Y}) - \log(\text{X}/\text{Y})_{\odot}$. [X/Fe] values are derived from empirical library abundance patterns and a response function as described in \citet{conroy18}. Select best-fit population parameters are listed in Table~\ref{tab:alf}. The age is a mass-weighted combination of the dominant and young stellar components. In the bottom half of the table we include the [O/H] and $\log{\epsilon_{\text{O}}} = 12 + \log(\text{N}_{\text{O}}/\text{N}_{\text{H}})$ oxygen abundances. \texttt{alf} adopts \citet{asplund}'s solar abundances: $\log{\epsilon_{\text{C}}} = 8.43$, $\log{\epsilon_{\text{N}}} = 7.83$, $\log{\epsilon_{\text{O}}} = 8.69$, and $\log{\epsilon_{\text{Mg}}} = 7.60$. 

\begin{table}
\caption{Stellar population and abundance parameters for the 24'' wide long slit spectra as best fit by \texttt{alf}. Ages are mass-weighted based on a two component, dominant and young, stellar age model. Abundance errors are set to 0.05 dex, the model systematic error, if statistical errors are less. $[\text{X}/\text{Y}] = \log(\text{X}/\text{Y})- \log(\text{X}/\text{Y})_{\odot}$ with solar abundances from \citet{asplund}. Stellar oxygen abundances are included in the bottom half of of the table.}
\label{tab:alf}
\centering
\begin{tabular}{lccc}
    \hline \hline
     & NGC 2768 & NGC 3245 & NGC 4694\\
    \hline
    age (Gyr)& 12.88 $\pm$ 1.60 & 10.55 $\pm$ 1.34 & 2.95 $\pm$ 0.56 \\
    $[\text{Fe}/\text{H}]$ & -0.10 $\pm$ 0.05 & -0.07 $\pm$ 0.05 & 0.22 $\pm$ 0.05 \\
    $[\text{C}/\text{Fe}]$ & 0.12 $\pm$ 0.05 & 0.12 $\pm$ 0.05 & -0.15 $\pm$ 0.05 \\
    $[\text{N}/\text{Fe}]$ & 0.10 $\pm$ 0.05 & 0.10 $\pm$ 0.05 & -0.16 $\pm$ 0.11 \\
    $[\text{O}/\text{Fe}]$ & 0.05 $\pm$ 0.05 & 0.13 $\pm$ 0.05 & 0.33 $\pm$ 0.22 \\
    $[\text{Mg}/\text{Fe}]$ & 0.18 $\pm$ 0.05 & 0.15 $\pm$ 0.05 & 0.38 $\pm$ 0.06 \\
    \hline
    $[\text{O}/\text{H}]$ & -0.06 $\pm$ 0.05 & 0.06 $\pm$ 0.05 & 0.55 $\pm$ 0.22 \\
    $12 + \log(\text{O}/\text{H})$ & 8.63 $\pm$ 0.05 & 8.75 $\pm$ 0.05 & 9.24 $\pm$ 0.22 \\
    \hline
\end{tabular}
\end{table}

\begin{figure*}
 \includegraphics[width=\textwidth]{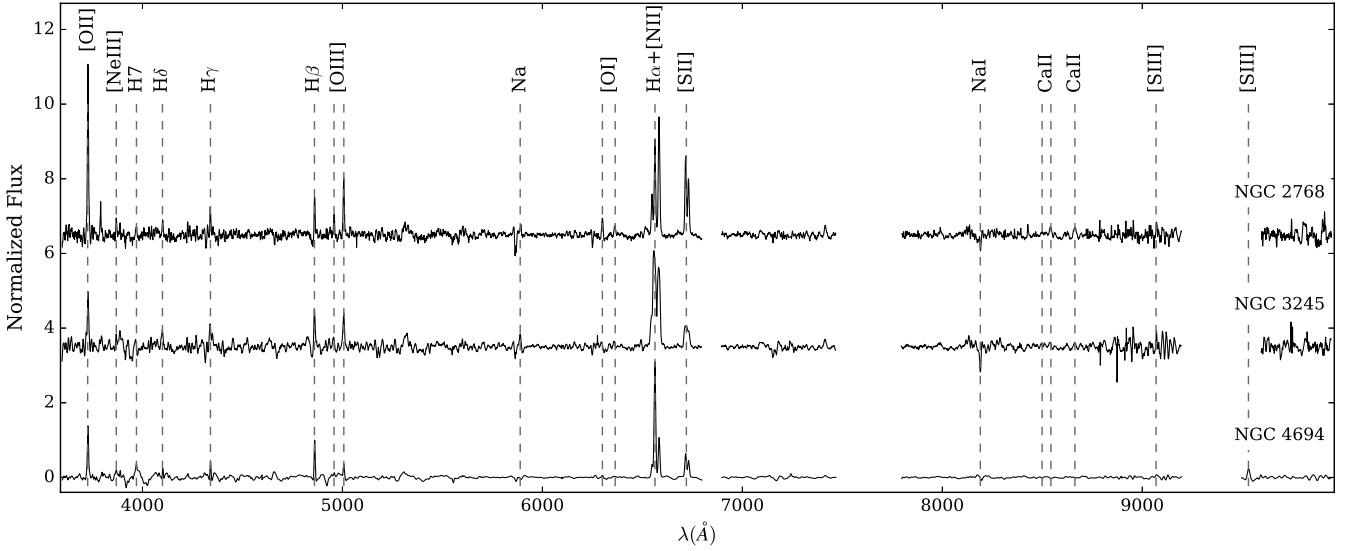}
 \caption{Emission line spectra of NGC 2768, NGC 3245, and NGC 4694. These spectra were normalized such that the peak of $\text{H}\beta \, \lambda 4861 = 1$, and then the spectra for NGC 3245 and NGC 2768 were shifted by 3.5 and 6.5, respectively. Spectra are shifted to rest-frame wavelengths and prominent lines are labeled. The telluric A and B bands, as well as the noisy region near 9400\AA, have been masked. All spectra are a summation over the central 24''.}
 \label{fig:spectras}
\end{figure*}

Though the fits shown in Figure~\ref{fig:alf_fit} include emission line features, a subsequent \texttt{alf} routine uses the fit parameters to produce solely stellar spectral models. These \texttt{alf} models are of higher resolution than our observed spectra. We convolve the stellar models with Gaussians to match the MODS instrumental resolution. We then continuum-normalize our observed spectra and convert to \texttt{alf}'s flux units. Finally, we subtract the model stellar spectra from the observed spectra, leaving the emission line spectra of our three objects (Figure~\ref{fig:spectras}).

\subsection{Stellar Masses and Star Formation Rates}
\label{sec:mass}

We calculate stellar masses using WISE W1 magnitudes and the M/L relationship derived by \citet{simonian}. We choose this relationship because that study also focused on dust. Our stellar masses agree with available S4G values and the results of similar calculations from \citet{cluver}. 

We also derive star formation rates (SFRs) for our sample with the infrared (IR) and ultra-violet UV star formation relationships from \citetalias{leroy}, who consolidate the work of \citet{kennicutt} and \citet{jarrett}. Star formation can be traced by UV light emitted by the young stellar population and IR light as dust particles absorb UV photons and re-emit them at redder wavelengths. Due to the dusty nature of these galaxies, we expect to see evidence of star formation in both the IR and UV. Therefore, we derive SFRs using a combined IR and UV indicator. We take IR and FUV/NUV magnitudes from the WISE and GALEX archives, respectively. Both W4 (22 $\mu$m) and W3 (12 $\mu$m) bands trace SF efficiently. W4 provides the best SF signal but has low sensitivity and a bias towards luminous IR galaxies. W3 has greater sensitivity than W4 and can detect dimmer galaxies. Large galaxy samples often derive SFRs from W3 magnitudes as they better represent the population. We choose to use W3 as our IR indicator because we have included a large sample of galaxies from previous work. 

If no dust were present we would expect the NUV and FUV star formation indicators to be identical. We find that the NUV band contributes more to the total UV star formation indicator than the FUV band and therefore choose NUV as the UV indicator. We sum the IR and UV components to give the total SFR. The stellar masses and SFRs are listed in Table~\ref{tab:obs}. 

Table~\ref{tab:obs} also includes the objects' specific star formation rate (sSFR), where $\text{sSFRs} = \text{SFR}/\text{M}_{\odot}$. We find values of $1.4\times 10^{-12}$ yr$^{-1}$ for NGC 2768, $2.5\times 10^{-12}$ yr$^{-1}$ for NGC 3245, and $8.8\times 10^{-12}$ yr$^{-1}$ for NGC 4694. Circumstellar dust in old ETGs will mimic a sSFR of $2 \times 10^{-12}$ yr$^{-1}$ \citep{simonian}, so W3 and W4 fluxes can not be trusted as SFR indicators below this threshold. NGC 2768 falls below this value, NGC 3245 near it, and NGC 4694 above it. Further, we expect that we overestimate these SFRs due to the old stellar population. Old stars within galaxies emit UV light and heat dust grains. This inflates the IR and UV luminosities and thus the measured SFR \citep{catalan}. As we assumed the UV and MIR luminosities are due to star formation alone, yet the sSFRs are close to or below the \citet{simonian} threshold, the star formation rates we adopt may be upper limits. All objects do show evidence of ongoing SF based on far-infrared emission \citep{martini13}. 

\subsection{Gas-Phase Abundances}
\label{sec:abundance}

To derive the emission line fluxes, we fit spectral features with Gaussian profiles. We use a non-linear, least squares minimization routine to fit the lines and surrounding continuum. We allow for freedom in line height, width, and a small range of central wavelength. The triple line system of $[\ion{N}{ii}]\lambda 6547$, H$\alpha\lambda 6562$, and $[\ion{N}{ii}]\lambda 6583$ is fit with three Gaussians of fixed width. Flux values ascertained from these Gaussian fits agree well with simple summations. We derive line flux errors from the continuum RMS.

We do not detect the auroral lines necessary for the direct method. Instead, we employ lines commonly used in strong line methods, $[\ion{O}{ii}]\lambda3727$, $\text{H}\beta\lambda4861$, $[\ion{O}{iii}]\lambda5007$, $\text{H}\alpha\lambda6563$, and $[\ion{N}{II}]\lambda6583$, with the following diagnostic emission line ratios:
\begin{flalign*}
    \text{N2} &=  \text{[\ion{N}{ii}]}\lambda6583/ \text{H}\alpha &\\
    \text{O3N2} &= \text{[\ion{O}{iii}]}\lambda 5007/\text{H}\beta/\text{[\ion{N}{ii}]} \lambda 6583/\text{H}\alpha &\\
    \text{N2O2} &= \text{[\ion{N}{ii}]}\lambda 6583/\text{[\ion{O}{ii}]} \lambda 3727 &
\end{flalign*}
These strong line indicators provide the gas-phase oxygen abundance, $ \log \epsilon_{\text{O}}= 12+ \log(\text{N}_{\text{O}}/\text{N}_{\text{H}})$.

Given the dusty nature of these galaxies, we expect that dust reprocessing has shifted some blue light to redder wavelengths. We quantify the amount of reprocessing through the Balmer decrement, where we measure the $\text{H}\alpha$ and $\text{H}\beta$ values with Gaussian fits. The line ratio should intrinsically exhibit the Case B recombination value of $\text{H}\alpha$/$\text{H} \beta = 2.87$ at $T_e = 10^4$K. We de-redden the spectra under this assumption according to the extinction law from \citet{cardelli}. The N2 and O3N2 indicators are relatively insensitive to reddening as they employ adjacent lines. The top row of Table~\ref{tab:metal} lists the $\text{H}\alpha/\text{H}\beta$ ratios prior to internal reddening corrections. The second section contains de-reddened flux values for the features used in the subsequent strong-line analysis. While \texttt{alf} also calculates emission line fluxes, it does not account for reddening or blended features. We choose to use our flux measurements rather than those fit by alf.

\begin{table}
\caption{Line ratios and gas-phase abundance results. Top row: $\text{H}\alpha/\text{H}\beta$ ratio after the Galactic reddening correction, but prior to correction for internal reddening. 
Second section: Reddening corrected line fluxes normalized to $\text{H}\beta=1$. Line fluxes  are based on a summation of the flux in the central 24'' along the slit. Errors are derived from the continuum RMS. 
Third section: Oxygen abundances derived from \citet{kd02} (KD02), \citet{pp04} (PP04), and \citet{pmc09} (PMC09). Values have the uncertainty of their calibrations: 0.18 dex for PP04 N2, 0.21 dex for PMC09 N2, 0.16 dex for PP04 O3N2, and 0.05 dex for KD02 N2O2. 
Bottom section: Oxygen abundances re-calibrated for SFR dependence \citep{brown}. A $\Delta\log(\text{sSFR})=-0.25$ was adopted for each galaxy. All values have a calibration uncertainty of 0.2 dex.}
\label{tab:metal}
\centering
\begin{tabular}{lccc}
    \hline \hline
    Ion/Indicator & NGC 2768 & NGC 3245 & NGC 4694\\
    \hline
    $\text{H}\alpha/\text{H}\beta$ & 3.54 $\pm$ 0.21 & 5.82 $\pm$ 0.46 & 4.62 $\pm$ 0.22 \\
    \hline
    $[\ion{O}{ii}]$ $\lambda3727$ & 6.15 $\pm$ 0.38 & 3.54 $\pm$ 0.33 & 2.86 $\pm$ 0.16 \\
    $\text{H}\beta$ $\lambda4861$ & 1.00 $\pm$ 0.08 & 1.00 $\pm$ 0.11 & 1.00 $\pm$ 0.07 \\
    $[\ion{O}{iii}]$ $\lambda5007$ & 1.62 $\pm$ 0.13 & 0.97 $\pm$ 0.10 & 0.33 $\pm$ 0.04 \\
    $[\ion{O}{i}]$ $\lambda6300$ & 0.46 $\pm$ 0.07 & 0.08 $\pm$ 0.03 & 0.04 $\pm$ 0.01 \\
    $\text{H}\alpha$ $\lambda6563$ & 2.86 $\pm$ 0.17 & 2.86 $\pm$ 0.23 & 2.86 $\pm$ 0.13 \\
    $[\ion{N}{II}]$ $\lambda6583$ & 3.57 $\pm$ 0.21 & 2.34 $\pm$ 0.19 & 0.97 $\pm$ 0.05 \\
    \hline
    PP04 N2 & 8.95 & 8.85 & 8.63 \\
    PMC09 N2 & 9.15 & 9.00 & 8.70 \\
    PP04 O3N2 & 8.69 & 8.71 & 8.73 \\
    KD02 N2O2 & 9.03 & 9.05 & 8.92 \\
    \hline
    Brown N2 & 9.22 & 9.12 & 8.89 \\
    Brown O3N2 & 8.99 & 9.00 & 9.03 \\
    Brown N2O2 & 9.16 & 9.19 & 9.04 \\
    \hline
\end{tabular}
\end{table}

NGC 2768 presents a unique reddening correction challenge due to the impact of atmospheric dispersion on the data. Instead of aligning the slit with the galaxy's parallactic angle, the slit was oriented with respect to the molecular gas features (Table~\ref{tab:obs}). Compounding this slit position with low hour angle observations caused a significant amount of blue light loss. We study the evolution of the $\text{H}\alpha/\text{H}\beta$ ratio over images of progressing hour angles. The changing value confirms that atmospheric dispersion impacts our data. While the dispersion should evenly scatter the smoothly distributed starlight in and out of the slit, we do not necessarily recover photons from \ion{H}{II} regions due to their non-uniform distribution throughout the galaxy. Given the effects of atmospheric dispersion, we cannot reliably determine NGC 2768's reddening and the atmospheric dispersion correction. In our analysis we will put more weight on indicators which employ adjacent lines because they are least sensitive to the effects of dust reprocessing and atmospheric dispersion. Observations of NGC 3245 and NGC 4694 were oriented along the galaxies' parallactic angles. 

The most common strong-line diagnostics include N2, O3N2, and N2O2, defined above. \citet{brown} assesses the performance of these indicators in a study of $\sim$200,000 star-forming galaxies from the Sloan Digital Sky Survey. They conclude that O3N2 is the most accurate indicator, followed closely by N2O2. They note N2O2's good performance at high oxygen abundance, an echo of Kewley \& Dopita (2002) who show that this indicator has a much tighter correlation with true abundances for objects with $12 + \log(\text{O}/\text{H}) > 8.6$. The N2 indicator under performs O3N2 and N2O2 as it worsens at higher oxygen abundance.  Overall, however, \citet{brown} concluded that all three relationships fair well and none excel far over the others. \citet{bresolin} also compare direct and strong-line indicators. In a spatially resolved study of M83, they find that all strong-line methods produce shallower gradients than the direct method. They also find PP04 O3N2 and KD02 N2O2 indicators lie systematically above the direct method abundances, though accounting for dust depletion may correct this discrepancy. We continue in our analysis with the N2, O3N2, and N2O2 indicators, adopting oxygen abundance definitions from \citet{pp04} (PP04), \citet{pmc09} (PMCO9), and \citet{kd02} (KD02), as well as the new calibrations from \citet{brown}. The third section of Table~\ref{tab:metal} contains the abundance results for our galaxies.

Oxygen abundances are also dependent upon the star formation rate of a galaxy \citep{ellison,lara,mannucci}. Recent work by \citet{brown} re-calibrates the above diagnostics against direct method abundances while accounting for the SFR dependence of the mass-metallicity relationship.  Their new metrics depend on $\Delta\log(\text{sSFR}) = \log(\text{sSFR}) - \langle\log(\text{sSFR})\rangle_{\text{M}_{*}}$ \citep{salim}, where $\langle\log(\text{sSFR})\rangle_{\text{M}_{*}}$ is the median $\log(\text{sSFR})$ of galaxies at a given $\text{M}_*$. We recalculate our oxygen abundance indicators according to the N2, N2O2, and O3N2 calibrations specified in \citet{brown}. We also calculate these indicators for the data in \citet{athey09} (9 ETGs observed with the Michigan-Dartmouth-MTT 2.4 m Hiltner Telescope), \citet{annibali} (57 ETGs observed with the 1.5 m ESO-La Silla telescope), and \citet{bresolin13} (NGC 404 observed by the Gemini Multi-Object Spectrograph at the Gemini North telescope). 

We find that ETGs in our sample and the literature fall near or beyond the high mass limit of \citet{brown}'s population and far below the $\Delta\log(\text{sSFR})$ lower limit. We do not extrapolate the relationships to such low $\Delta\log(\text{sSFR})$ as there is little to no data in that regime. Instead, we choose to assign all galaxies a $\Delta\log(\text{sSFR})=-0.25$, the lowest bin included in \citet{brown} and the closest approximation to our sample. We know that these galaxies SF, so some correction will improve our oxygen abundances. It is reasonable to adopt this higher value of $\Delta\log(\text{sSFR})$ because the MZR's dependence on sSFR lessens in higher mass galaxies, especially for the O3N2 indicator. These abundance values are listed in the bottom section of Table~\ref{tab:metal}. If we were to use the \citet{brown} values for the actual $\Delta\log(\text{sSFR})$, all indicators would show much higher oxygen abundances. 

\subsection{Alternative Ionization Sources}
\label{sec:ionization}

Gas-phase oxygen abundances such as KD02 and PP04 are calibrated through abundance measurements of extragalactic \ion{H}{ii} regions and photoionization models. However, ETGs contain many other ionization sources, such as nuclear activity, shocks, cosmic rays, magnetohydrodynamic waves, and old, hot stars. We evaluate the potential impact of each of these ionization sources on our abundance measurements.

First, we assess the star-forming nature of our objects and their potential AGN contamination by placing them on the classic Baldwin-Phillips-Terevich (BPT) diagram \citep{baldwin}. BPT diagrams typically include [\ion{O}{iii}]$\lambda5007$, as this state has a higher ionization energy than other prominent strong-lines. The hardness of the radiation field sets the [\ion{O}{iii}]$\lambda5007$/H$\beta$ ratio. A high value implies more energetic photons and suggests AGN activity. We determine our galaxies' principle energy and ionization sources by locating them in $\log([\text{\ion{O}{iii}}]\lambda5007/\text{H}\beta)$ vs. $\log([\text{\ion{N}{ii}}]\lambda6583/\text{H}\alpha)$ space and comparing their position to star-formation/AGN boundaries, such as those found by \citet{kauff} and \citet{kewley06}. \citet{kauff} derive their SF/AGN boundary by examining a sample of ~60,000 emission line galaxies. Within this sample, a clear SF and LINER sequence emerge. \citet{kewley06} and \citet{kewley01} use theoretical models to establish an extreme starburst line--an upper limit on the emission line strengths in star-forming galaxies. Star formation dominates galaxies which fall below these two curves, while those above may be classified as LINERs or AGNs. If a galaxy falls within the AGN locus, we expect that its off-nuclear line ratios would lie closer to the star-forming region.

\begin{figure*}
 \includegraphics[width=\textwidth]{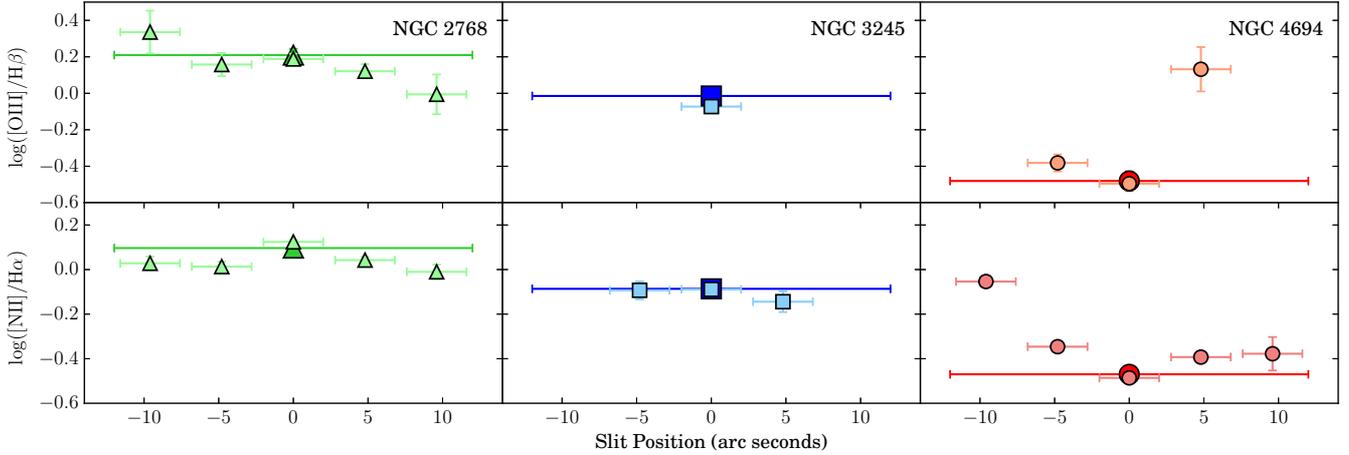}
 \caption{Ratio of [\ion{O}{III}]$\lambda5007$/H$\beta$ and [\ion{N}{II}]$\lambda6583$/H$\alpha$ for different slit positions and the entire galaxy slit for NGC 2768 (left), NGC 3245 (middle) and NGC 4694 (right). All line flux values are de-reddened. The bars in the x-dimension represent the region summed along the slit. The bars in the y-dimension represent the uncertainty on the ratio values. When not visible, error bars are smaller than the markers. $\text{H}\beta$ was not detected in all spectra and the [\ion{O}{III}]$\lambda5007$/H$\beta$ values have been excluded in these cases.}
 \label{fig:ratio}
\end{figure*}

In addition to spectra of the entire galaxy (extraction width 24"), we extract multiple smaller, one-dimensional spectra from each galaxy: one on the nucleus and two to four off-nuclear positions of width 4". We apply the analysis described in Section~\ref{sec:alf} and \ref{sec:abundance} to each. Figure~\ref{fig:ratio} shows the [\ion{O}{iii}]$\lambda5007$/H$\beta$ and [\ion{N}{ii}]$\lambda6583$/H$\alpha$ ratio along the spatial dimension (lighter, smaller markers), as well as the values of the entire slit (darker, larger markers). $\text{H}\beta$ was not detected in some off-nuclear spectra. We find that the off-nuclear line flux measurements concur with nuclear and full galaxy line fluxes in NGC 2768 and NGC 3245. NGC 4694 displays larger line ratio fluctuations in off nuclear spectra. The change in emission line strength causes the derived oxygen abundance to vary by 0.2-0.3 dex across the slit. These off nuclear oxygen abundances are within 1-2$\sigma$ of the nuclear and full slit values. Due to the consistency in oxygen abundance across the slit, the remainder of our analysis employs just the full galaxy spectra.

\begin{figure}
 \includegraphics[width=\columnwidth]{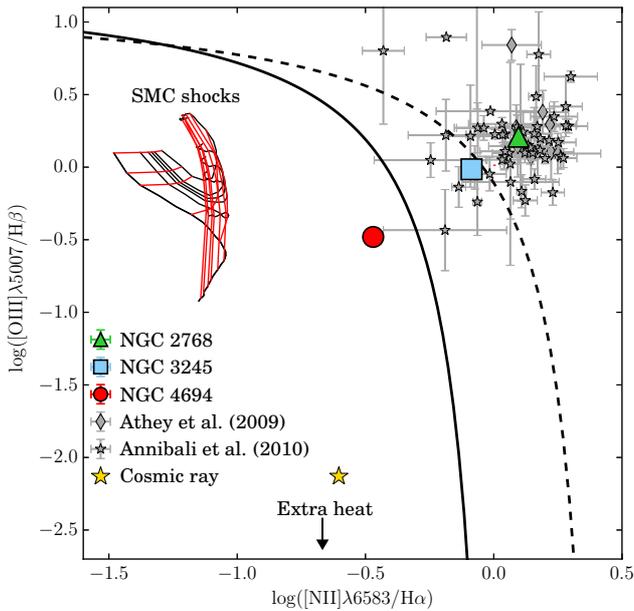}
 \caption{BPT diagram including star formation/AGN boundaries from \citet{kauff} (solid line) and \citet{kewley06} (dashed line). Error bars on data from this work are smaller then the markers. We also mark the locations of alternate ionization sources (the extra heat point falls off of the plotted region). The SMC shock models represent constant magnetic parameter (0.5-10 $\mu$G cm$^{3/2}$, left to right black lines) and constant shock velocity (125-1000 km s$^{-1}$, top to bottom red lines). The cosmic ray and extra heat line ratios are from \citet{ferland} and the SMC shock models are from the MAPPINGS III library \citep{allen}. See Section~\ref{sec:ionization} for further details.}
 \label{fig:bpt}
\end{figure}

Figure~\ref{fig:bpt} shows our galaxies on a BPT diagram. NGC 4694 falls below both SF/AGN boundaries and is clearly in the star formation locus. NGC 3245 is close to the extreme star formation limit line of \citet{kewley06}. NGC 2768 is above both curves and therefore in the LINER region, along with ETGs from previous studies by \citet{athey09} and \citet{annibali} (grey markers). We applied reddening corrections and calculated oxygen abundances for galaxies from the literature as we did for our sample (see Section~\ref{sec:abundance}). The line ratios for NGC 2768, and the galaxies in \citet{athey09} and \citet{annibali}, suggest that there is AGN contamination in the $[\ion{N}{ii}]\lambda6583$ line. As mentioned in Section~\ref{sec:methods},  we expect some AGN activity in NGC 2768. 

If ionization is coming from nuclear activity, we expect that off nuclear spectra would show lower emission line ratios. NGC 2768's off-nuclear spectra show slightly lower $[\ion{N}{ii}]\lambda6583/\text{H}\alpha$ ratios but similar $[\ion{O}{iii}]\lambda5007/\text{H}\beta$ values (Figure~\ref{fig:ratio}). These points would also fall in the LINER region of the BPT diagram. NGC 3245's off-nuclear [\ion{N}{ii}]$\lambda6583$/H$\alpha$ values are similarly high. Though both have nuclear line ratios indicative of AGN, neither have the lower off-nuclear emission line ratios characteristic of star formation. We rule out AGN ionization in NGC 2768 and NGC 3245 because of the line flux ratios are consistent across the entire galaxy.

Alternatively, emission from old, hot stars could explain NGC 2768 and NGC 3245's positions on the BPT diagram. Post-asymptotic giant branch (pAGB) stars are known to be sources of ionizing photons \citep{binette}, especially among old stellar populations such as those seen in ETGs. Integrated field studies have convincingly shown that the line ratios and line fluxes of some galaxies do not decrease with radius as rapidly as expected if nuclear activity were the only way to produce LINER-like ratios \citep{sarzi}. \citet{belfiore} studied the location of many star forming galaxies on the BPT diagram and found that they can lie near or beyond the SF/LINER threshold, due to pAGB ionization. They estimate that with average stellar mass surface densities pAGB star emission could be of the same order of magnitude as weak AGN. They conclude that many galaxies dominated by star formation are misclassified as active because of this overlap. We refer the reader to their paper for a more detailed discussion of pAGB ionization.

In addition to pAGB stars, ionization by fast or slow shocks can influence the observed line fluxes. Previous work by \citet{crocker} and \citet{sarzi} suggested that the line ratios in NGC 2768 could be affected by shock ionization in the central region. While their SAURON project used [\ion{O}{iii}]$\lambda5007$/H$\beta$ and [\ion{N}{i}]$\lambda\lambda5197,5200/\text{H}\beta$ ratios as diagnostic tools, their wavelength coverage does not extended to redder lines such as [\ion{N}{ii}]$\lambda6583$. We use the MAPPINGS III shock models \citep{allen} to determine the potential effect of shock ionization on the [\ion{N}{ii}]/H$\alpha$ ratio. The \citet{veilleux} log([\ion{O}{iii}]$\lambda5007$/H$\beta$) vs. log([\ion{N}{ii}]$\lambda6583$/H$\alpha$) diagnostic diagrams place NGC 2768 on the edge of low velocity, solar shock models. Similar diagnostic diagrams for different atomic abundances show large variations between low metallicity shock models and super-solar shock models \citep[Figure 21]{allen}. The shock model for an SMC like galaxy (low metallicity) with velocities of 125-1000 km s$^{-1}$ and magnetic parameters of 0.5-10 $\mu$G cm$^{3/2}$ is included in Figure~\ref{fig:bpt}.  This low metallicity shock model occupies a much lower log([\ion{N}{ii}$\lambda6583$]/H$\alpha$) space than our objects. Our objects are consistent with shock models for MW-metallicity gas, but they are inconsistent with shocks in low-metallicity (SMC-like) gas. The MAPPINGS III models demonstrate that shocks cannot make sub-solar metallicity gas appear solar or super-solar.

Collisional heating from cosmic rays and dissipative magnetohydrodynamic waves (``extra heat'') could also contribute to low-ionization emission line fluxes. \citet{ferland} modeled spectra which reproduced H$_2$/H$\alpha$ ratios for both extra heating and cosmic ray cases to within a factor of two. We take line fluxes from their work and calculate the [\ion{O}{iii}]$\lambda5007$/H$\beta$ and [\ion{N}{ii}]$\lambda6583$/H$\alpha$ ratios characteristic of these two ionization mechanisms. We include these values on the BPT diagram in Figure~\ref{fig:bpt}--though the extra heat case falls off of the plotted region. Given that both models occupy regions not populated by our galaxies, we  rule out extra heating and cosmic rays as potential ionization sources.  

\begin{figure}
 \includegraphics[width=\columnwidth]{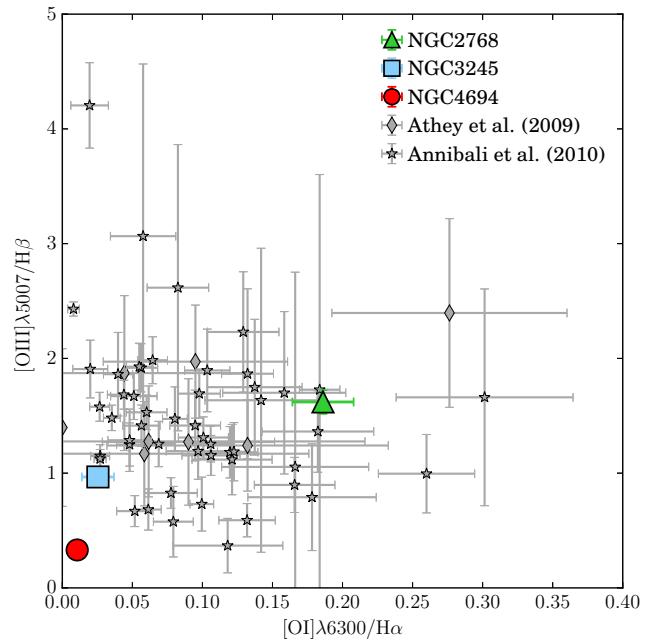}
 \caption{Diagnostic diagram from \citet{woods} for ionization by the stellar population with a contribution from SD progenitors. Systems dominated by SD ionization have $[\ion{O}{I}]/\text{H}\alpha > 0.5$ and would fall outside of this plot. Error bars on data from this work are smaller than the markers for some cases.}
 \label{fig:woods}
\end{figure}

Finally we consider ionization by single degenerate (SD) Type Ia supernovae progenitors \citep{woods}. In this scenario a white dwarf (WD) accretes matter from its main sequence or red giant companion until the WD explodes as a supernovae. If this formation theory is correct and SD supernovae occur at the expected rate, then the accreting WD population should be extremely luminous. In young populations ($<4$ Gyrs) of temperatures $10^5-10^6$K, SD progenitors could be the dominant ionization source. To determine if WDs play an active role in ionizing gas, \citet{woods} develop diagnostic tools which utilize $[\ion{N}{I}]$, $[\ion{O}{I}$], and $[\ion{C}{II}]$--emission lines enhanced through high temperature ionization. We reproduce their $[\ion{O}{I}]\lambda6300/\text{H}\alpha$ vs. $[\ion{O}{III}]\lambda5007/\text{H}\beta$ diagnostic diagram in Figure~\ref{fig:woods}. Their SD models predict $[\ion{O}{I}]\lambda6300/\text{H}\alpha \geq \sim 0.5$ for $\text{T}_{\text{eff}}=10^6$K and solar gas-phase abundances--beyond the plotted region. Our sample, as well as that of \citet{athey09} and \citet{annibali}, have emission line ratios below this threshold and thus do not show evidence for a contribution from hot SD progenitors. \citet{woods} include a sub-sample of points from \citet{annibali} and come to a similar conclusion, though they do not rule out the SD case due to uncertainties in the line fluxes. Our line fluxes have lower uncertainties and our results are correspondingly less ambiguous. 

We conclude that the ionized gas in NGC 4694 is due to photionization by young stars, while there is some contribution from old stars in NGC 3245 and these stars may dominate in NGC 2768. We do not find any evidence that AGN, cosmic ray heating, extra heat, shocks, or hot SD progenitors significantly affect the line ratios. The main result of this analysis is that we find no evidence for physical processes that could make low-metallicity gas from accreted dwarf galaxies look like the higher-metallicity gas that we expect from internal production. While pAGB stars may contribution to the ionization of the gas, this contribution should not make the ionized gas appear to have a substantially higher metallicity.

\section{Results}
\label{sec:results}

\subsection{Comparing Stellar and Gas-Phase Oxygen Abundances}
\label{sec:stellar_ab}

\begin{figure*}
 \includegraphics[width=\textwidth]{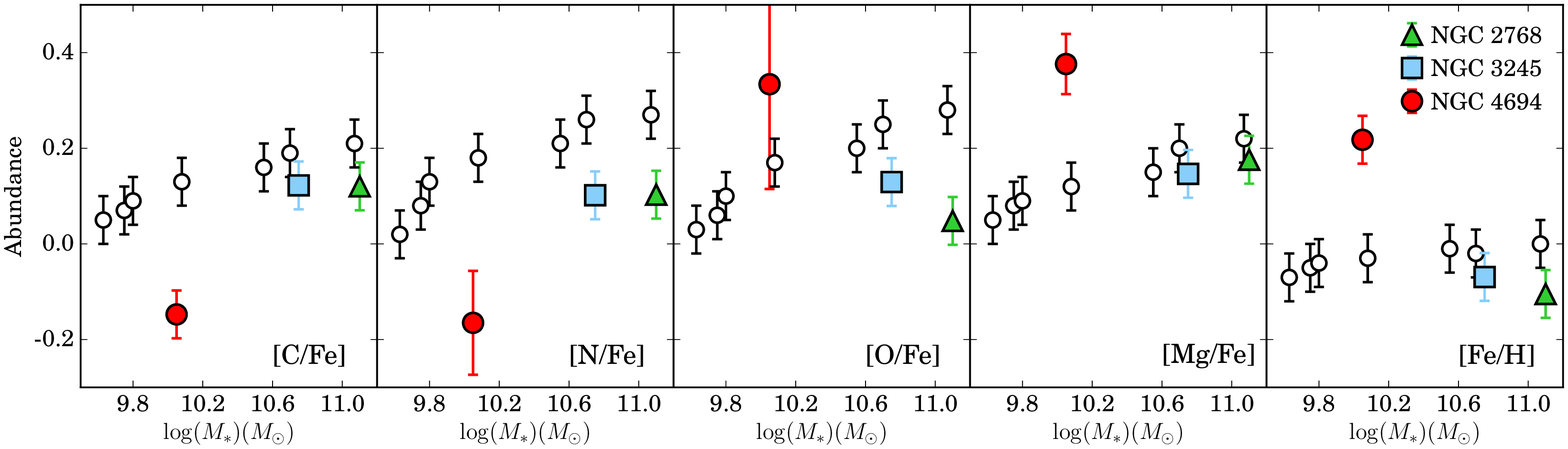}
 \caption{A comparison of \texttt{alf}'s best-fit $[\text{C}/\text{Fe}]$, $[\text{N}/\text{Fe}]$, $[\text{O}/\text{Fe}]$, $[\text{Mg}/\text{Fe}]$, and $[\text{Fe}/\text{H}]$ values to ETG abundance trends from \citet{conroy14}. \texttt{alf} measurements and the trends adopt solar oxygen abundances from \citet{asplund}.}
 \label{fig:c_metals}
\end{figure*}

We compare the stellar and gas-phase oxygen abundances to determine if the ionized gas is consistent with internal production or external accretion. The main stellar abundances we consider are [Fe/H], [C/Fe], [N/Fe], [O/Fe], and [Mg/Fe]. In a study of stacks of SDSS ETGs binned by velocity dispersion, \citet{conroy14} found that N, C, and Mg trace O, as expected from models of chemical evolution. We compare our best-fit stellar abundances to their measurements in Figure~\ref{fig:c_metals} where we have plotted their results vs. stellar mass, rather than velocity dispersion. The model stellar abundances also use solar ratios from \citet{asplund} and include 0.05 dex of systematic errors \citep{conroy14}. We adopt this minimum uncertainty on our stellar abundances in all further analysis. Most abundance values for NGC 2768 and NGC 3245 lie within 1-2$\sigma$ the ETG stellar abundance trends found by \citet{conroy14}. NGC 4694's abundances differ by a larger amount, outside the realm of statistical errors. These differences are likely due to NGC 4694's relatively young stellar population (Table~\ref{tab:alf}), which sets it apart from a typical ETG. At such young ages the abundance sensitivity decreases and abundances are more difficult to measure. The disagreement between some of our data and the models could be due to abundance gradients within the galaxies. In spiral galaxies, we expect the uncertainty due to this effect to be $\sim 0.1-0.2$ dex \citep{kudritzki14,bresolin}. ATLAS$^{\text{3D}}$ and SDSS-IV MANGA have found radial gradients in ETG Lick indices ($\sim 0.2-0.4$ \AA) and metallicity ($\sim 0.1-0.2$ dex) as well \citep{scott,sdss}.   

\begin{figure}
 \includegraphics[width=\columnwidth]{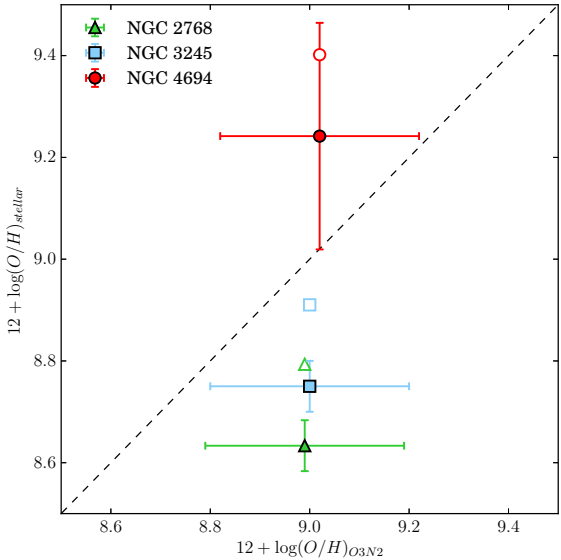}
 \caption{A direct comparison of stellar and gas-phase oxygen abundances. We take the Brown O3N2 values as the best gas-phase abundance. Solid symbols adopt solar oxygen values from \citet{asplund} while hollow symbols adopt solar oxygen values from \citet{villante}. }
 \label{fig:comp_metals}
\end{figure}

We directly compare the stellar and gas-phase oxygen abundances of our ETGs in Figure~\ref{fig:comp_metals}. We take stellar oxygen abundances from \texttt{alf} (Table~\ref{tab:alf}) and take \citet{brown}'s O3N2 calibrations to be the best gas-phase oxygen abundance. A direct comparison of these data is shown by the solid markers. While \texttt{alf} adopts a solar oxygen abundance of $\log\epsilon_{\text{O}}=8.69$ \citep{asplund}, we note that recent work by \citet{villante} have found a higher $\log\epsilon_{\text{O}}=8.85$ from solar seismology. If we adopt this value, the stellar oxygen abundance would increase by 0.16 dex to the hollow markers in Figure~\ref{fig:comp_metals}. We find that for both solar abundance values, the gas-phase and stellar oxygen abundances are consistent within 1-2$\sigma$.

\begin{figure}
 \includegraphics[width=\columnwidth]{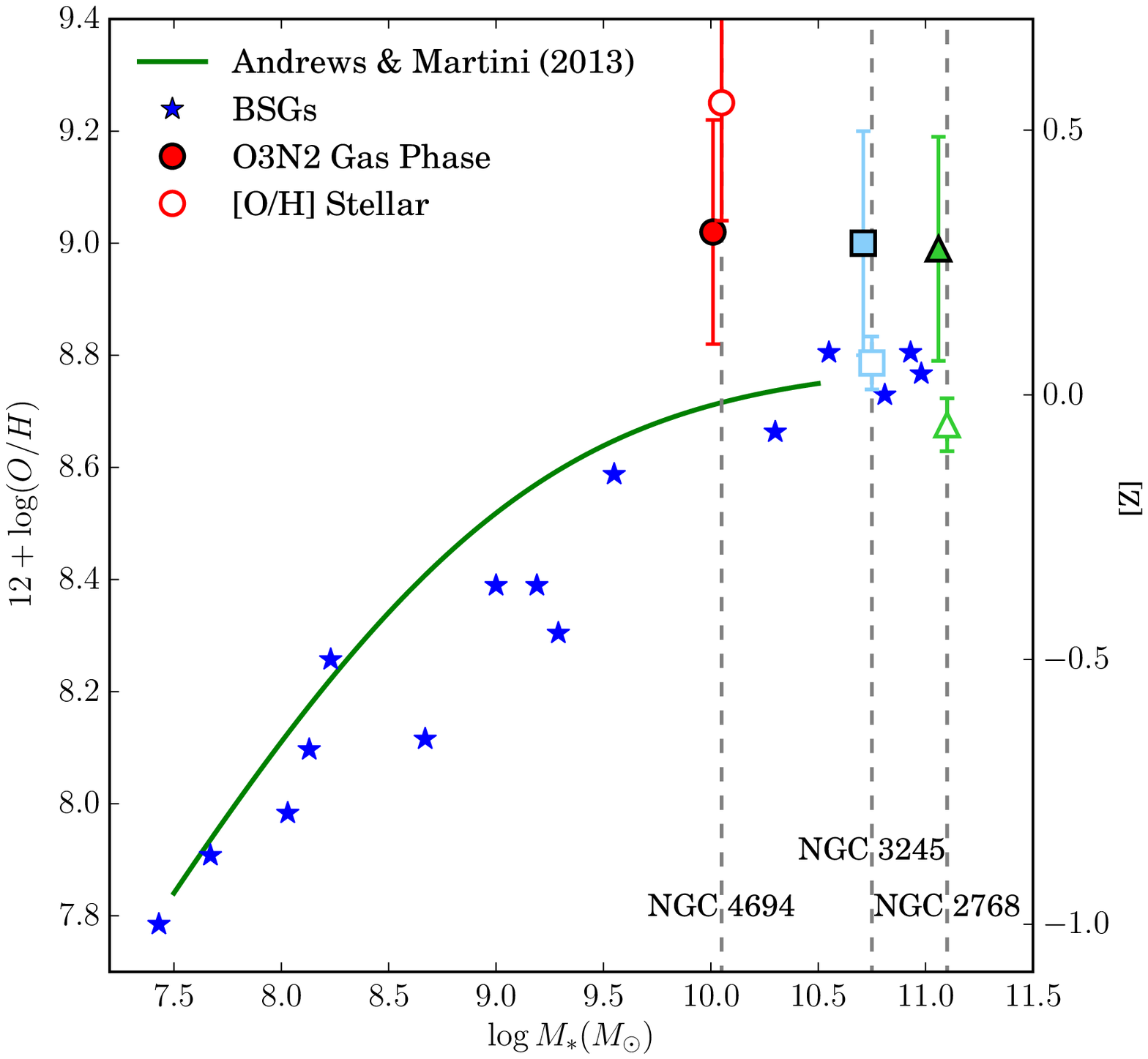}
 \caption{Stellar [O/H] (open symbols) and Brown O3N2 gas-phase oxygen abundances (solid symbols) compared to \citet{andrews} MZR (green line) and BSGs from \citet{kudritzki14}, \citet{hosek}, \citet{kudritzki16} and \citet{bresolin} (blue stars). Gas-phase abundances have been offset by -0.04 $\log{\text{M}_*(\text{M}_\odot)}$ for clarity.}
 \label{fig:bresolin}
\end{figure}

Finally, we compare our stellar and gas-phase abundances to work by \citet{bresolin} on stellar abundances for nearby galaxies from blue supergiants (BSGs) and the mass-metallicity relationship. Figure~\ref{fig:bresolin} shows the gas-phase O3N2 abundance values for each galaxy based on \citet{brown} (solid symbols), the best-fit [O/H] stellar abundances (open symbols), stellar abundances derived from BSGs in \citet{kudritzki14}, \citet{hosek}, \citet{kudritzki16}, and \citet{bresolin}, and the mass-metallicity relation from \citet{andrews}. All stellar abundances in this plot adopt the \citet{asplund} solar oxygen value. The stellar abundances agree with the nebular abundance trends, as well as the measured gas-phase oxygen abundances for their respective galaxies. Stellar oxygen abundances for NGC 2768 and NGC 3245 follow the BSG trend. NGC 4694's stellar abundance falls above the BSG trend due to the galaxy's younger age, as discussed above.

Based on the results shown in Figures~\ref{fig:comp_metals} and \ref{fig:bresolin}, we do not find evidence of gas-phase oxygen abundances that are significantly lower than that of the stellar population. This observation argues strongly against external accretion of low metallicity dwarf galaxies as the source of dust and gas in ETGs. 

\subsection{The Mass-Metallicity Relationship}
\label{sec:mzr}

\begin{figure*}
 \includegraphics[width=\textwidth]{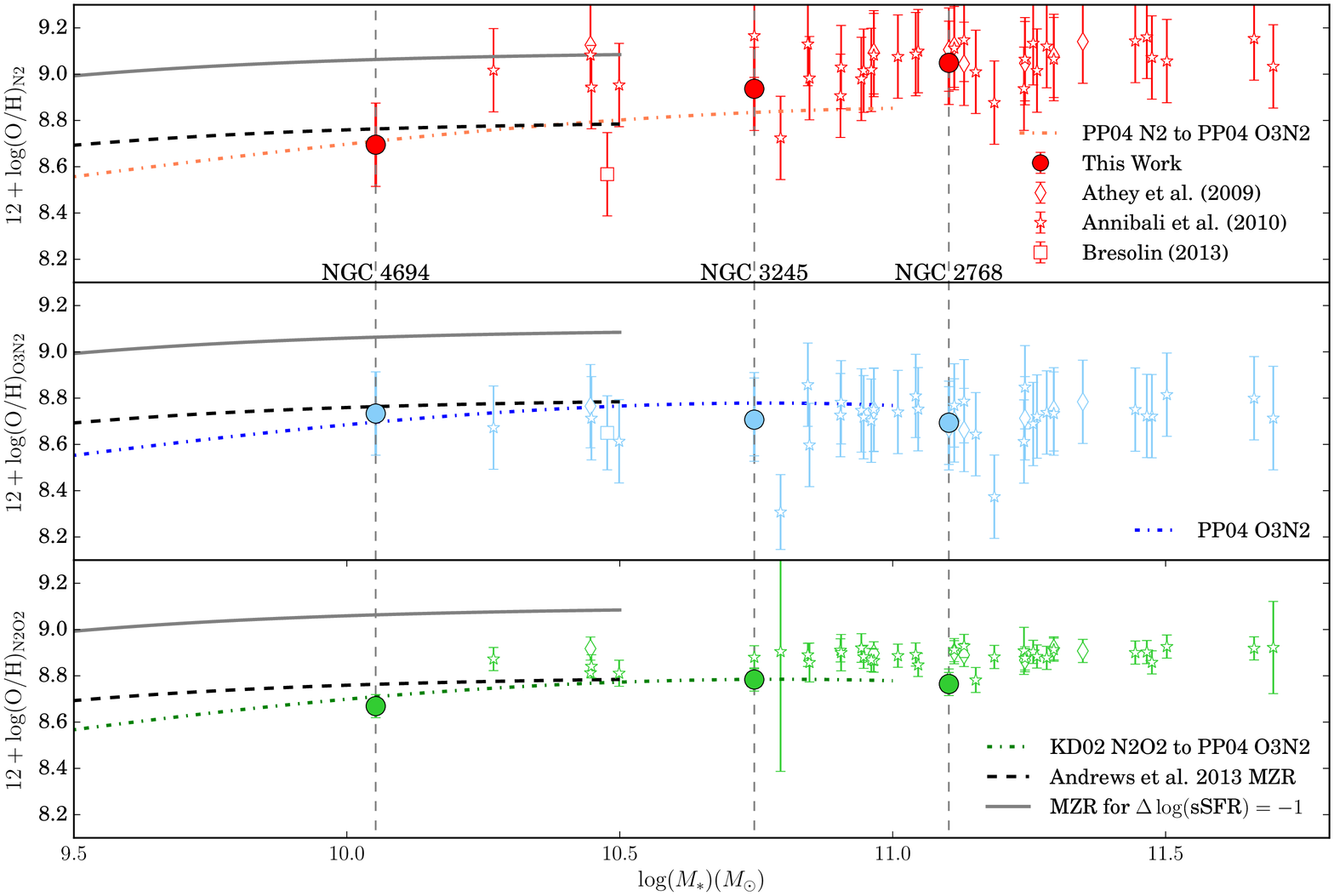}
 \caption{MZR for PP04 N2 (top), PP04 O3N2 (middle), KD02 N202 (bottom) indicators. All strong-line MZRs and ETG abundances have been re-calibrated to the PP04 O3N2 metric for comparison. Points include oxygen abundance measurements of NGC 2768, NGC 3245, and NGC 4694 from this paper (circles), \citet{athey09} (diamonds), \citet{annibali} (stars), and \citet{bresolin13} (square). 
Some error bars from this work are smaller than the markers. MZRs are taken from \citet{kewley08} and \citet{andrews}. The MZR for $\Delta\log(\text{sSFR})=-1$ shows how the \citet{andrews} MZR would shift for low $\Delta\log(\text{sSFR})$ galaxies, such as ours, if we included star formation according to the \citet{brown} prescription.}
 \label{fig:mzr}
\end{figure*}

We compare our metallicity determinations to the mass-metallicity relation (MZR) \citep{tremonti04,kewley08,andrews} in Figure~\ref{fig:mzr}. The location of our sample in mass-metallicity space with respect to theoretical and/or empirical MZRs is a separate way to determine how their gas-phase abundances relate to a typical galaxy of similar mass. Figure~\ref{fig:mzr} plots the PP04 N2 (top), PP04 O3N2 (middle), and KD02 N2O2 (bottom) ETG abundances with MZR from \citet{kewley08} for the respective calibration. In order to compare the three mass-metallicity trends, we must use the same calibration. We convert all PP04 N2 and KD02 N2O2 ETG abundances and both MZRs to the PP04 O3N2 relationship \citep{kewley08}. We choose the PP04 O3N2 calibrator as its MZR shows good agreement with the direct method MZR in the high mass range \citep{andrews}. All data points in Figure~\ref{fig:mzr} fall near their respective models. The O3N2 and N2O2 indicators show tighter relationships than the N2 indicator. We note that many of the N2 points fall outside of the PP04 O3N2 re-calibration range ($12+\log{[\text{O/H}]}=8.05-8.8$) and that they are an extrapolation of the relationship. Figure~\ref{fig:mzr} includes a sub-sample of points from \citet{athey09} and \citet{annibali} for which the specified abundances and masses could be derived. \cite{bresolin13}'s N2 and O3N2 measurements of NGC 404 with a mass from \citet{thilker} is also included. In all three relationships, none of our objects show signatures of low metallicity gas.

The mass-metallicity relationship is correlated with the star formation rate \citep{ellison, lara, mannucci}. At fixed mass, galaxies with higher SFRs have lower metallicities than those with low SFRs. Figure~\ref{fig:mzr} includes the \citet{andrews} MZR, as well as a $\Delta\log(\text{sSFR})=-1$ MZR for low $\Delta\log(\text{sSFR})$ based on the slope of the $\Delta\log(\text{sSFR})$ oxygen abundance anti-correlation derived by \cite{brown}. This MZR increases the \citet{andrews} model by 0.3 dex. This model is an extrapolation, but we expect that galaxies with negative $\Delta\log(\text{sSFR})$, such as those in this sample, will have systematically higher oxygen abundances if SF is accounted for. We emphasize that there are limited data for gas-phase abundances at the high stellar masses we consider here. Even less data are available at this range of SFRs.

Of the three indicators, O3N2 agrees best with the models. Given the high oxygen abundances, we expect O3N2 to outperform N2 and N2O2 as gas cools more efficiently at higher temperatures and metallicities. Nitrogen becomes the dominant coolant at high metallicities, leading to lower flux [\ion{N}{II}] emission lines \citep{kd02}. \citet{brown} note that the N2 indicator saturates at metallicities above solar. Further, O3N2 employs ratios of more adjacent lines than N2O2, making it less susceptible to the effects of atmospheric dispersion. We therefore put more weight on the O3N2 metric.

Our MZR (Figure~\ref{fig:mzr}) shows that these three galaxies possess gas of solar or super solar oxygen abundance. This is also true for other ETGs with data from the literature. All three methods show consistent metallicities. No galaxy displays evidence of low metallicity gas. As the $\Delta\log(\text{sSFR})=-1$ MZR falls above the \citet{andrews} MZR, we can see that SF calibrations will act to increase a galaxy's metallicity. Oxygen abundances derived from \citet{brown} SF-dependent calibrations would lie above classic strong-line abundances seen in Figure~\ref{fig:mzr}. 

\section{Discussion}
\label{sec:discussion}

One hypothesis to explain the dust and gas in ETGs is minor mergers with low metallicity dwarf galaxies. Alternatively, the dust and gas could be internally produced by mass loss from evolved stars or cooling from the hot phase of the ISM. The relative uniformity of ETG stellar populations, in contrast to the presence of dust and cold gas in only $\sim$60\% of the population, is at odds with the internal production hypothesis.

ATLAS$^{\text{3D}}$ kinematics and dust morphology data also support external accretion and provide clues to the merger/interaction history of ETGs. In a study of galaxies with the CARMA ATLAS$^{\text{3D}}$ CO survey, \citet{alatalo} classify the disturbance of ETGs based on their molecular gas kinematics and morphology. Two of our three objects, NGC 2768 and NGC 4694, appear in their work and were placed into the disc/ring and extremely disrupted classes, respectively. \citet{alatalo} concluded that NGC 4694 belongs to an interacting system which caused its irregular CO distribution and kinematics. These results agree with previous work by \citet{duc} who propose VCC 2062 as the donor tidal dwarf galaxy. Interestingly, they find VCC 2062 has an oxygen abundance of $12 + \log(\text{O}/\text{H})=8.6-8.7$, approximately solar.  \citet{crocker} and \citet{alatalo} also discuss the asymmetric morphology of NGC 2768. They both conclude that the significant misalignnent of the kinematic major axes of the ionized gas and stars suggests NGC 2768 has undergone or is currently involved in a minor merger or other accretion event. Though the neighborhood consists of many potential donor galaxies, \citet{crocker} suggest UGC 4808 as the most likely companion. 

While there is some circumstantial evidence of minor mergers in these two cases, we do not measure the lower oxygen abundances that would be indicative of external accretion. More generally, the minor merger hypothesis requires a very high merger rate. Based on equations by \citet{stewart}, \citet{martini13} calculate a merger rate of $0.07-0.2$ Gyr$^{-1}$. Coupled with the short dust destruction timescale, $\tau_{\text{dust}} \sim 2\times10^4$yr \citep{draine_b}, they predict a dusty ETG fraction of $f_{\text{dust}} < 0.0014-0.004$ from mergers alone. Minor mergers can not produce the observed dust mass without a significantly higher merger rate or a significantly longer dust destruction timescale.

If minor mergers are the main source of gas and cold dust in ETGs, one might expect gas dilution to cause little change in the overall gas-phase oxygen abundance. This is not the case. While dusty ETGs contain some cold gas, they do not have the surplus necessary to dilute accreted material beyond detection. Instead, gas from the dwarf galaxy will dilute that of the ETG, lowering the overall metallicity. This post-interaction metallicity drop has been observed for merging systems. \citet{kewley06a} and \citet{ellison08} compare the MZRs of galaxy pairs and field galaxies. They find that the pairs have systematically lower metallicities, as interactions trigger gas flows which redistribute metal poor gas.  We expect interacting ETG to show similar metallicity trends. \citet{bresolin13} predict the post-merger gas-phase oxygen abundance of NGC 404 from \ion{H}{i} observations \citep{rio} and the star formation history \citep{williams}. They conclude that even after 0.6 Gyr of enhanced star formation, the gas-phase oxygen abundance of the ETG will remain sub-stellar by tenths of a dex. Our solar gas-phase abundances are inconsistent with the gas dilution and metal depletion expected from the merger scenario. 

Our oxygen abundance measurements are in excellent agreement with internal production of dust and gas from evolved stars or cooling from the hot phase of the ISM. While this hypothesis also struggles to produce the observed dust masses and the fraction of dusty ETGs, an ISM hot-phase temperature of $10^5$K or rapid cooling could lower the dust destruction timescales enough to bridge the dust mass gap \citep{martini13,mathews}. \citet{pulido} use CO and X-ray emission from ETGs to link molecular gas content and short cooling timescales. Their data agree better with an internal rather than external origin and they consequently conclude that the molecular gas condenses from the hot atmosphere. They propose that X-ray bubbles from the central AGN lift gas away from the nucleus and cause molecular gas to condense and cool as it falls inward. Recent work by \citet{babyk} finds similar trends for a small sample of early spiral and eliptical galaxies. X-ray measurements of our sample and other ETGs with and without cold gas and dust would be valuable to determine if there is a correlation between shorter cooling times and the presence of cold gas and dust.

\section{Summary}
\label{sec:summary}

We have presented long-slit spectroscopy of three dusty early-type galaxies with MODS on the LBT and measured their gas-phase and stellar abundances. Our main results are: 

(1) The emission line flux ratios do not vary significantly along the slit, so the oxygen abundances derived from the integrated spectra will be representative of the entire galaxy and not just the nucleus.

(2) The emission lines are predominantly ionized due to stars (either young or pAGB) and are not subject to other ionization sources that may impact the abundance measurements. NGC 2768 and NGC 3245 fall near the LINER region while NGC 4694 falls on the SF sequence. We expect some signatures of nuclear activity in NGC 2768, but the off-nuclear line ratios also fall within the LINER region, which points to ionization by the old stellar population. 

(3) We rule out alternative ionization by cosmic rays and extra heating due to the models' occupation of a different BPT parameter space than our objects. MAPPINGS III shock models suggest that solar metallicity shocks may be present in NGC 2768. We determine that these shocks will not disguise low metallicity gas as solar or super-solar gas. We conclude that our derived abundances reflect the true values.

(4) We derive gas-phase oxygen abundances from the N2, O3N2, and N2O2 strong-line indicators and calibrate these results to account for their SFR dependence. We derive stellar oxygen abundances through \texttt{alf}'s stellar population modeling. We find that the gas-phase and stellar oxygen abundances are in good agreement (Figures~\ref{fig:comp_metals} and \ref{fig:bresolin}). Additionally, all objects' gas-phase oxygen abundances lie along the MZR (Figure~\ref{fig:mzr}) and do not show the lower values expected from interactions or mergers with low-metallicity dwarf satellites.

(5) Although the irregular dust and gas morphology and kinematics suggests a merger origin, our results clearly show that the gas and stellar abundances are consistent. This strongly supports internal production by the old stellar population and/or cooling from the ISM rather than external accretion. Future X-ray observations to measure the cooling from the hot phase of the ISM would be valuable to investigate why only $\sim$60\% of otherwise similar ETGs have gas and dust.

\section*{Acknowledgments}

We thank Kevin Croxall, Michael Fausnaugh, Eric Huff, Roberta Humphrys, Christopher Kochanek, Rick Pogge, Suk Sien Tie, and Steven Villanueva for their contributions to the OSU/RC queue when these data were obtained.

This paper used data obtained with the MODS spectrographs built with funding from NSF grant AST-9987045 and the NSF Telescope System Instrumentation Program (TSIP), with additional funds from the Ohio Board of Regents and the Ohio State University Office of Research.

This paper made use of the modsIDL spectral data reduction reduction pipeline developed in part with funds provided by NSF Grant AST-1108693.

\texttt{alf} has been supported by Packard and Sloan foundation fellowships, NASA grants NNX14AR86G and NNX15AK14G and NSF grant AST-1524161. 

\bibliographystyle{mnras}
\bibliography{citations}

\bsp	
\label{lastpage}
\end{document}